\documentclass[usenatbib]{mn2e}
\pdfoutput=1
\usepackage{graphicx}

\newcommand{\beq}{\begin{equation}}
\newcommand{\eeq}{\end{equation}}
\newcommand{\beqary}{\begin{eqnarray}}
\newcommand{\eeqary}{\end{eqnarray}}
\newcommand{\lxpwn}{L_{\rm X,pwn}}
\newcommand{\lxpsr}{L_{\rm X,psr}}
\newcommand{\gmpwn}{\Gamma_{\rm pwn}}
\newcommand{\gmpsr}{\Gamma_{\rm psr}}
\newcommand{\bs}{B_{\rm s}}
\newcommand{\blc}{B_{\rm lc}}
\newcommand{\edot}{\dot{E}}
\newcommand{\pdot}{\dot{P}}
\newcommand{\Fxpwn}{F_{\rm X,pwn}}
\newcommand{\Fxpsr}{F_{\rm X,psr}}

\title[X-rays from PWNe and their PSRs]{The power-law component of the X-ray emissions from pulsar wind nebulae and their pulsars}
\author[Hsiang \& Chang]{Jr-Yue Hsiang,$^{1}$ Hsiang-Kuang Chang$^{1,2,3}$ \thanks{E-mail: hkchang@mx.nthu.edu.tw}
\\
$^{1}$Institute of Astronomy, National Tsing Hua University, Hsinchu 30013, Taiwan
\\
$^{2}$Department of Physics, National Tsing Hua University, Hsinchu 30013, Taiwan
\\
$^{3}$Center for Informatics and Computation in Astronomy (CICA), National Tsing Hua University, Hsinchu 30013, Taiwan
}

\begin{document}

\label{firstpage}
\date{Accepted 2021 January 4. Received 2021 January 4; in original form 2020 September 17}
\pubyear{2020}
\pagerange{\pageref{firstpage}--\pageref{lastpage}}
\maketitle

\begin{abstract}
To look for possible phenomenological connections between pulsar's timing properties and emissions from pulsar wind nebulae and their pulsars,
we studied 
the power-law component of the X-ray emissions
 from 35 pulsar wind nebulae which have a detected pulsar in X-rays.
Our major results are in the following:
(1) 
The power-law component of the 
X-ray luminosities, in the energy range from 0.5 keV to 8 keV, 
of the nebulae and of the pulsar both show a strong correlation with the pulsar spin-down power
 ($\edot$),
consistent with earlier studies.
However, equally significant  correlations with the 
magnetic field strength at the light cylinder  ($\blc$)
are also found.
The similar significance level of the correlations with $\edot$ and with  $\blc$
suggests that not only $\edot$ but also $\blc$ plays an important role in understanding these
power-law 
emissions. 
(2) Thermal X-ray emissions are detected in 12 pulsars among the 35 samples. 
With derived temperature as one additional variable,  
we found that the photon indices of pulsar's non-thermal X-ray power-law spectra can be well described
by  a linear function of
$\log P$, $\log\pdot$ and temperature logarithm $\log T$.
It indicates that the surface temperature of neutron stars plays an important role in determining
the energy distribution of the radiating pair plasma in pulsar's magnetospheres. 
\end{abstract}


\section{Introduction}

Pulsar wind nebulae (PWN) are observed across the electromagnetic spectrum 
from radio bands to very high energy gamma rays. 
They are extended sources powered by the embedded pulsar through its magnetized winds.
These winds are energetic charged particles, most likely electron-position pairs, flowing out from 
pulsar's magnetosphere. When they encounter the relatively slow-moving supernova ejecta, or the interstellar medium 
if the supernova remnant is no longer around, a pulsar wind nebula shock and a reverse one, called
the termination shock, are formed (e.g. \citet{slane17}). 
It is generally believed that the wind  is Poynting-dominated when just leaving the magnetosphere,
that is, its energy density is dominated by the Poynting flux rather than by particles.
The wind, however, becomes particle-dominated when approaching the termination shock
\citep{kennel84}. 
It was suggested that, based on a one-zone, shocked wind model, the nebula X-ray luminosity $\lxpwn$
depends on the central pulsar's spin-down power $\edot$ as $\lxpwn\propto \edot^{\alpha/2}$, 
where $- \alpha$ is the power index of the power-law energy distribution of the emitting particles \citep{chevalier00,cheng04}.
We note that, however, mechanisms 
of accelerating  wind particles are a subject of investigation.
It was argued that Fermi acceleration at the termination shock is not able to produce particles energetic enough to 
explain the observed synchrotron radiation in X-ray bands and beyond.  Instead, 
magnetic reconnection way ahead of or around the termination shock was proposed to be the mechanism
of converting magnetic energy to particles' energy \citep{lyubarsky03,sironi13}.
On the other hand, based on the outer-gap model,
non-thermal X-ray luminosity of pulsars, $\lxpsr$, was predicted to be proportional to
pulsar's spin-down power roughly as $\lxpsr\propto\edot^{1.15}$ with some variations due to different magnetic inclination angles \citep{cheng99,cheng04}.
It is therefore of interest to compare observations of X-ray luminosities from the nebula and from pulsar's magnetosphere
with these model predictions.

Luminosity of X-ray emissions 
 ($L_{\rm X}$) 
 from pulsars, sometimes mixed with their PWN, has been indeed found to 
strongly correlate with pulsar's spin-down power $\edot$.
In \citet{seward88}, with {\it Einstein} data of nine pulsars, 
it was found that $L_{\rm X}$ (0.2 -- 4 keV) is proportional to $\edot^{1.39}$, where,
the luminosity $L_{\rm X}$ is contributed from both the pulsar and their nebulae. 
With {\it ROSAT} data of 27 pulsars, \citet{becker97} reported that $L_{\rm X}$ (0.1 -- 2.4 keV) is about $10^{-3}\edot$.
This luminosity counts only the emissions from the pulsar.
To reduce the contribution from surface thermal emissions, \citet{possenti02} considered the energy range of
2 -- 10 keV and reported that $L_{\rm X} \propto \dot{E}^{1.34}$, with a sample of 39 pulsars,
whose PWNe, if present, were not separated from their pulsars. 
Therefore in that study the luminosity includes contribution from PWNe if they exist.
Another relation of $L_{\rm X} \propto \dot{E}^{1.87}$ (2 -- 10 keV) was reported in
\citet{mattana09} for 14 TeV-emitting PWNe, some of which are mixed with pulsar's luminosity.

In the works mentioned above, there seems to be an indication that the relation between the X-ray luminosity $L_X$ and 
the spin-down power $\edot$ has a larger power index when emissions from PWNe are included
and a smaller one when only the emission from the pulsar is considered.
Indeed, in \citet{cheng04}, with a sample of 23 pulsars in the energy range between 2 and 10 keV,
the non-pulsed luminosity was found to be $L_{\rm X}\propto \dot{E}^{1.4}$,
while the pulsed one was $L_{\rm X}\propto \dot{E}^{1.2}$.
In \citet{li08}, with  24 PWNe and 27 pulsars, it was reported that, in 2 -- 10 keV,
$L_{\rm X,pwn}\propto \dot{E}^{1.45}$ and $L_{\rm X,psr} \propto \dot{E}^{0.92}$,
where $\lxpwn$ is the luminosity from PWNe and $\lxpsr$ is that from pulsars.
It is one of the purposes of this paper to examine these relations with a larger sample.

The spectral photon indices $\gmpwn$ and $\gmpsr$ in the power-law spectra of 
the non-thermal X-ray emissions 
from PWNe and from pulsars are another important 
property of the emission. 
With nine bright Crab-like pulsars, relations between $\gmpwn$ and $\gmpsr$, between $\gmpwn$ and $\edot$,
and between $\gmpsr$ and $\dot{E}$ were reported \citep{gotthelf03}.
These relations were, however, not confirmed in \citet{li08} with a larger sample of pulsars and their PWNe.
Nonetheless, \citet{cheng04} reported an indication of a positive correlation  between $\gmpwn$ and $\lxpwn/\dot{E}$,
and  \citet{li08} found moderately strong positive correlations
between $\gmpwn$ and $\lxpwn$ and between $\gmpwn$ and $\lxpwn/\dot{E}$.
On the other hand,
$\gmpsr$ was not found to correlate with  luminosities or other pulsar timing properties \citep{cheng04,li08}.

In this paper, we report correlations among properties of PWN/pulsar 
power-law component of the 
X-ray emissions and of pulsars' rotation, using a sample of 35 PWNe, all with detected pulsars whose X-ray emission can be separated from that of PWNe.
To have a more uniform comparison basis, we consider 
the  flux
 in the energy range from 0.5 keV to 8 keV
for all the sources in the sample. This flux is derived from the power-law component  of the corresponding
best-fit spectral models reported in the literature. It therefore does not include surface thermal emissions
(i.e. also excluding the emission ultimately originating from bombardment of electron-positron pairs produced and accelerated in the magnetosphere, see the next section).

Thanks to observations with fine enough spatial resolution and long enough exposure, 
the X-ray emission from PWNe has been found to become softer with increasing distance from the pulsar,
that is, the photon index is larger at a larger distance. Clear examples can be found in 
\citet{slane12} and \citet{pavan16}. That certainly represents the cooling of emitting particles in PWNe and deserves
detailed study by its own. We therefore decided not to include $\gmpwn$ in our study,
and to keep only $\lxpwn$, $\lxpsr$, and $\gmpsr$ as spectral properties of sources in our sample.
On the other hand, in studying these possible correlations, we also take the magnetic field strength at the light cylinder and
the surface temperature (when available) into account. 
In particular, the inclusion of the surface temperature
leads to a statistically much better description of $\gmpsr$ 
as a linear function of $\log P$, $\log \dot{P}$ and $\log T$.

We describe, in Section 2, how we collect the sample and determine their properties from the literature.
In Section 3, results of possible correlations among spectral and timing properties are presented, for
the 35 sources in the sample as a whole, 
and for a subset of 12 samples whose surface temperature estimates are available.
Implications of these results are discussed in Section 4.
     
\section{Sample collection}
\begin{table*}
\renewcommand{\arraystretch}{1.55}
\caption{X-ray spectral properties of 35 PWNe with detected pulsars. 
References: (1) \citet{slane04}; (2) \citet{li08}; (3) \citet{klingler16b}; (4) \citet{ge19}; (5) \citet{deluca05}; (6) \citet{posselt17};
(7) \citet{birzan16}; (8) \citet{pavlov01}; (9) \citet{kargaltsev07}; (10) \citet{pavan16}; (11) \citet{chang12}; (12) \citet{ng05};
(13) \citet{kishishita12}; (14) \citet{klingler16a}; (15) \citet{delaney06}; (16) \citet{kargaltsev09}; (17) \citet{romani05};
(18) \citet{hinton07}; (19) \citet{klingler18}; (20) \citet{roberts03}; (21) \citet{helfand07} ; (22) \citet{duvidovich19}; (23) \citet{matheson10}; 
(24) \citet{gotthelf08}; (25) \citet{petre02}; (26) \citet{temim10}; (27) \citet{becker06}; (28) \citet{li05}; (29) \citet{vanetten08};
(30)  \citet{romani17}; (31) \citet{johnson10}.
$^\dagger$PSR J2124-3358 is a millisecond pulsar, which is not included in the selected sample of 35 objects to which
the correlation analysis was applied,
but is listed here for reference.
}
\begin{tabular}{lllllll}
\hline
\hline
Pulsar name & PWN name  & $\lxpwn$ (erg/s)   & $\lxpsr$ (erg/s)  & $\gmpsr$   & $kT_{\infty}$ (eV)   & References \\
\hline
J0205+6449  &G130.7+3.1 &$2.26^{+2.22}_{-1.46}\times 10^{34}$& $1.25^{+1.33}_{-0.83} \times 10^{34}$
   & $2.07^{+0.02}_{-0.02}$ & $112.0^{+8.6}_{-8.6}$    & 1, 2\\
J0358+5413 &  & $5.89^{+6.00}_{-3.83}\times 10^{31}$& $7.38^{+8.05}_{-4.90}\times 10^{30}$ 
  & $1.45^{+0.21}_{-0.24}$ & $160.0^{+12.0}_{-12.0}$   & 3 \\
J0534+2200  &G184.6-5.8& $1.89^{+1.99}_{-1.24}\times 10^{37}$& $1.32^{+1.89}_{-0.97}\times 10^{36}$ 
   & $1.63^{+0.09}_{-0.09}$ & --   & 2 \\
J0537-6910  &G279.6-31.7  & $2.09^{+2.60}_{-1.45}\times 10^{36}$& $7.43^{+14.00}_{-5.54}\times 10^{35}$ 
  & $1.80^{+0.20}_{-0.20}$ & -- &  2 \\
J0540-6919   &G279.7-31.5  & $1.43^{+1.38}_{-0.92}\times 10^{37}$& $3.02^{+3.36}_{-2.01}\times 10^{36}$ 
    & $0.78^{+0.09}_{-0.09}$   & --      & 4 \\
J0633+1746  &G195.1+4.3  & $3.37^{+3.49}_{-2.21}\times 10^{29}$& $1.48^{+1.46}_{-1.02}\times 10^{30}$ 
  & $1.47^{+0.06}_{-0.07}$ &$83.4^{+4.2}_{-4.2}$  & 5, 6 \\
J0659+1414 & & $8.35^{+19.30}_{-6.94}\times 10^{28}$& $1.51^{+2.83}_{-1.07}\times 10^{30}$ 
   & $2.30^{+0.68}_{-0.57}$ & $93.1^{+4.6}_{-7.3}$   & 7 \\
J0835-4510 &G263.9-3.3 & $7.14^{+7.76}_{-4.74}\times 10^{32}$& $2.20^{+2.70}_{-1.52}\times 10^{32}$ 
   & $2.20^{+0.40}_{-0.40}$ & $130.1^{+2.6}_{-2.6}$    & 2, 8 \\
J1016-5857& G284.3-1.8 & $2.00^{+2.89}_{-1.46} \times 10^{32}$         
   & $5.00^{+2.00}_{-2.00} \times 10^{31}$& $1.50^{+0.40}_{-0.40}$       & --    & 9 \\
J1048-5832  & G287.4+0.58 & $6.60^{+7.52}_{-4.44} \times 10^{31}$             & $2.60^{+0.40}_{-0.40} \times 10^{31}$& $1.50^{+0.30}_{-0.30}$     & --                             & 9 \\
J1101-6101 &IGRJ11014-6103 & $7.43^{+7.37}_{-4.80}\times 10^{33}$& $3.53^{+3.56}_{-2.29}\times 10^{33}$ 
   & $1.08^{+0.08}_{-0.08}$ & --                      & 10 \\
J1124-5916 &G292.0+1.8 & $3.14^{+5.01}_{-2.29}\times 10^{34}$& $2.66^{+4.09}_{-1.91}\times 10^{33}$ 
    & $1.62^{+0.10}_{-0.10}$ & --                   & 2 \\
J1357-6429  && $2.24^{+3.10}_{-1.60}\times 10^{32}$ &$7.27^{+14.90}_{-5.41}\times 10^{31}$
   &  $1.72^{+0.55}_{-0.63}$ & $140.0^{+60.0}_{-40.0}$ &  11 \\
J1420-6048   &G313.6+0.3& $3.39^{+3.41}_{-2.20}\times 10^{34}$& $4.63^{+8.62}_{-3.73}\times 10^{32}$ 
    & $1.00^{+4.20}_{-4.80}$ & --                   &  12, 13 \\
J1509-5850   &    & $1.22^{+1.23}_{-0.79}\times 10^{33}$& $7.14^{+5.51}_{-4.72}\times 10^{31}$ 
    & $1.90^{+0.12}_{-0.12}$ & --    &   14 \\
J1513-5908  & G320.4-1.2& $1.35^{+1.64}_{-0.93}\times 10^{35}$& $9.86^{+16.20}_{-7.54}\times 10^{34}$ 
   & $1.19^{+0.04}_{-0.04}$ & --                   &  15 \\
J1617-5055 & & $1.84^{+2.14}_{-1.25}\times 10^{33}$& $9.52^{+9.21}_{-6.10}\times 10^{33}$ 
    & $1.14^{+0.06}_{-0.06}$ & --             &  16 \\
J1709-4429 &G343.1-2.3 & $1.21^{+1.21}_{-0.78}\times 10^{33}$& $1.82^{+1.92}_{-1.19}\times 10^{32}$ 
    & $1.62^{+0.20}_{-0.20}$ & $172.3^{+14.6}_{-13.8}$&  17  \\
J1718-3825 &      & $2.93^{+5.68}_{-2.40} \times 10^{32}$& $2.43^{+4.72}_{-1.99} \times 10^{32}$ 
    & $1.47^{+0.21}_{-0.21}$ & -- & 18 \\
J1747-2958 &G359.23-0.82& $5.15^{+5.05}_{-3.32}\times 10^{33}$& $1.48^{+1.77}_{-1.01}\times 10^{33}$
   & $1.55^{+0.04}_{-0.04}$ & --        &  19  \\
J1801-2451 & G5.27-0.9 & $2.12^{+3.28}_{-1.58}\times 10^{32}$& $1.54^{+1.70}_{-1.03}\times 10^{33}$ 
   & $1.60^{+0.60}_{-0.50}$ & --                   &  9 \\
J1803-2137  && $2.00^{+2.30}_{-1.35}\times 10^{32}$ & $6.76^{+8.38}_{-4.67}\times 10^{31}$ 
   & $1.40^{+0.60}_{-0.60}$ & $200.0^{+100.0}_{-100.0}$&  9 \\
J1809-1917  && $8.60^{+10.20}_{-5.86}\times 10^{32}$& $3.23^{+5.15}_{-2.38}\times 10^{31}$ 
   & $1.23^{+0.62}_{-0.62}$ & $170.0^{+30.0}_{-30.0}$ & 19 \\
J1811-1925 &G11.2-0.3 & $1.70^{+1.71}_{-1.10}\times 10^{34}$& $6.97^{+7.27}_{-4.57}\times 10^{33}$ 
    & $0.97^{+0.32}_{-0.39}$ & --                  &   20 \\
J1813-1749 &G12.82-0.02 & $2.55^{+4.95}_{-2.09}\times 10^{34}$& $5.92^{+11.49}_{-4.85}\times 10^{33}$
    & $1.3^{+0.30}_{-0.30}$  & --                  &   21 \\
J1826-1334 &G18.0-0.7 & $2.09^{+2.13}_{-1.36}\times 10^{33}$& $1.78^{+1.89}_{-1.17}\times 10^{32}$ 
    & $1.26^{+0.25}_{-0.25}$ & --                  &   22 \\
J1833-1034 & G21.5-0.9 & $2.35^{+2.78}_{-1.60}\times 10^{34}$ & $9.05^{+12.20}_{-6.44}\times 10^{33}$ 
   & $1.14^{+0.05}_{-0.07}$ & $520.0^{+30.0}_{-40.0}$&  23 \\
J1838-0655 &     & $6.73^{+11.57}_{-5.25} \times 10^{32}$ & $3.55^{+3.92}_{-2.36} \times 10^{34}$
   &  $0.50^{+0.20}_{-0.20}$ & -- &  24 \\
J1856+0113 &G34.7-0.4 & $6.80^{+8.11}_{-4.63}\times 10^{32} $& $1.08^{+1.59}_{-0.78}\times 10^{32}$ 
    & $1.28^{+0.48}_{-0.48}$ & --                  &  25 \\
J1930+1852 & G54.1+0.3  & $5.31^{+10.30}_{-4.36}\times 10^{34}$& $1.55^{+3.00}_{-1.27}\times 10^{34}$ 
    & $1.20^{+0.20}_{-0.20}$ & --                  & 26 \\
J1932+1059 & G47.4-3.9 & $6.13^{+18.60}_{-4.84}\times 10^{29}$& $2.97^{+3.11}_{-1.96}\times 10^{30}$ 
   & $2.72^{+0.12}_{-0.09}$ & --                  &   27 \\
J1952+3252 & G69.0+2.7 & $7.86^{+8.35}_{-5.18}\times 10^{33}$& $2.91^{+3.28}_{-1.95}\times 10^{33}$ 
    & $1.63^{+0.03}_{-0.05}$ & $130.0^{+2.0}_{-2.0}$ &   28 \\
J2021+3651 & G75.2+0.1 & $6.82^{+7.11}_{-4.47}\times 10^{33}$& $6.34^{+7.33}_{-4.29}\times 10^{32}$ 
    & $1.73^{+1.15}_{-1.02}$ & $159.4^{+17.2}_{-20.6}$ &  29 \\
J2124-3358$^\dagger$ &  & $3.34^{+3.35}_{-2.16} \times 10^{29}$& $2.80^{+2.84}_{-1.82} \times 10^{29}$
    & $2.50^{+0.20}_{-0.10}$ & $250.0^{+10.0}_{-10.0}$  & 30 \\
J2225+6535 &G108.6+6.8 & $4.53^{+5.57}_{-3.25}\times 10^{30}$ & $1.45^{+1.73}_{-1.03}\times 10^{30}$ 
    & $1.70^{+0.46}_{-0.23}$ & --           &   31 \\
J2229+6114 & G106.6+3.1 & $4.04^{+7.83}_{-3.06}\times 10^{32}$ & $3.82^{+6.30}_{-2.70}\times 10^{32}$ 
   & $1.05^{+0.10}_{-0.10}$ & --                  &   2 \\
\hline
\end{tabular}
\label{tab:sp}
\end{table*}

We collected, from the literature, all the PWNe whose associated pulsars are detected.
In this effort, among many different references, we benefited very much from the McGill PWN catalog \footnote{Roberts, M.S.E., 2004, `The Pulsar Wind Nebula Catalog (March 2005 version)', McGill University, Montreal, Quebec, Canada (available on the World-Wide-Web at `http://www.physics.mcgill.ca/~pulsar/pwncat.html').} and \citet{li08}. 
Altogether we found 50, among which 35 have separate 
X-ray spectral information for PWNe and for their associated pulsars.
These 35 are the sources that we study in this work.
Their spectral properties are listed in Table \ref{tab:sp}. 

In several earlier studies \citep{possenti02,cheng04,li08} the X-ray luminosity considered 
was the total luminosity in the energy range of 2 -- 10 keV.
That luminosity includes contributions of emissions from the nebulae, emissions from pulsars' magnetosphere,
and likely also emissions from the stellar surface due to bombardment of charged particles flowing back from the
magnetosphere.
The luminosity of the bombardment component presumably may depend on pulsars' timing properties
in a way different from that of the magnetospheric emission.
We therefore try to separate them by considering only 
 the flux
derived from the best-fit power-law component in corresponding spectral analysis works to
represent the component of the magnetospheric emission. 
In fact, with the caveats above, hereafter we will always identify the power-law component of the X-ray spectra 
with the non-thermal emission and will use the two terminologies in an interchangeable way.
To
have a uniform comparison basis, we consider, for all the 35 sources, 
only the power-law component luminosities in the energy range from 0.5 keV to 8 keV.
This energy range is the most common one employed in the spectral analysis we found in the literature, 
in particular for those based on Chandra data.
When fluxes and uncertainties in the literature were reported in a different energy range, 
we compute the corresponding flux in 0.5 -- 8 keV and  estimate the uncertainty with the published value scaled by the flux ratio of the two energy ranges.
To covert fluxes into luminosities, distances to the sources are needed.
We adopt the best-estimate distances provided in the ATNF pulsar catalog\footnote{https://www.atnf.csiro.au/research/pulsar/psrcat/}
for the conversion. 
Following \citet{possenti02} and \citet{li08}, we also adopt an uncertainty of $\pm$40\% in distances.
For most of the cases, this is in fact the dominating uncertainty in determining luminosities.
These distances and other pulsar timing properties are listed in Table \ref{tab:tm}.
   
Among these 35 pulsars,  thermal emissions are detected in 12 of them, 
in addition to the non-thermal power-law component.  
These are thermal emissions from neutron stars' surface, which reveal information about the surface temperatures.
Because of the presence of strong magnetic fields and the anisotropy of thermal conductivity, 
the surface temperature is higher near magnetic poles. Furthermore, possible charged-particle bombardment
on the polar cap region heats the magnetic polar region further. 
Thermal emissions from the non-uniform surface temperature distribution is very often dominated
by the hot polar region. 
In spectral analysis, besides the power-law component, 
one usually employs a black-body component to describe the thermal emissions if that improves
the spectral fitting statistically.
Sometimes two black-body components, 
representing a hot spot at the magnetic pole and a cooler stellar surface outside the hot spot, 
are used when further improvement can be obtained. 
Some authors also use neutron-star atmospheric emission models to fit the thermal emissions.
Surface thermal emissions may affect the production of electron-positron pairs and their energy distribution,
either in the outer gap (e.g. \citet{cheng86,takata09}) or above the polar cap (e.g. \citet{chang95,sturner95,harding98,timokhin19}).
That in turn will affect spectral properties of non-thermal emissions from pulsars' magnetosphere.
Since the black-body model has been applied in all the 12 sources in the literature, 
we use the surface temperatures inferred from the best-fit black-body component for our correlation analysis.
For the cases of best fit with two black-body components (PSR J0633+1746 and PSR J0659+1414), 
a flux-weighted average temperature is employed. These temperatures are listed in Table \ref{tab:sp}
(in the `$kT_\infty$' column, where $T_\infty$ is the temperature inferred from the measured spectrum and $k$ is the
Boltzmann constant).

For the convenience of future investigation, pulsars with detected PWNe but not included in this study for 
various reasons are listed in the following:
(1) PWNe or pulsars too dim in X-rays to separate their X-ray emission  -- 
PSR J0437-4715 \citep{bogdanov13,guillot16,bogdanov19},
PSR J0538+2817 \citep{romani03,mcgowan03,ng07}, 
PSR J1028-5819 \citep{mignani12},
PSR J1057-5226  \citep{posselt15}, 
PSR J1301-6305 \citep{abramowski12},
PSR J1648-4611 \citep{sakai13},
PSR J1702-4128 \citep{chang08},
PSR J1831-0952 \citep{abichandani19};
(2) Pulsars {\bf with magnetar-like bursts} --
PSR J1119-6127 \citep{blumer17,lin18,dai18,archibald18}, 
PSR J1846-0258 \citep{kuiper18,reynolds18,temim19};
(3) A black-widow pulsar, whose X-ray emission is from an intrabinary shock --
PSR  J1959+2048 \citep{huang12,kandel19};
(4) An old millisecond pulsar appearing as an outlier in $P$ and $\pdot$ distributions -- 
PSR J2124-3358 \citep{romani17}; 
(5) No reported X-ray emissions --
PSR J0908-4913 \citep{gaensler98,stappers99},
PSR J1341-6220 \citep{wilson86,caswell92,kaspi92},
PSR J1646-4346 \citep{giacani01}.
More PWNe whose pulsars have not yet been detected can be found in the McGill PWN Catalog.
We note that, for the two pulsars with magnetar-like bursts, 
the X-ray luminosities, both of the nebulae and of the pulsars, were observed to vary \citep{blumer17,reynolds18}. 
Those emissions may come from a mixture of mechanisms powered by pulsar's rotation and strong magnetic fields.
PSR  J2124-3358 is a millisecond pulsar. Its non-thermal X-ray luminosity
 may depend on the timing properties
in a way different from normal pulsars, because of its weak magnetic field (e.g. \citet{zhang03}).
This pulsar is still included in all the figures for reference.

\begin{table*}
\renewcommand{\arraystretch}{1.38}
\caption{Properties of the 35 pulsars employed in our analysis. Surface temperature information is available for 12 of them.
$^\dagger$PSR J2124-3358 is a millisecond pulsar, which is not included in the selected sample of 35 objects to which
the correlation analysis was applied,
but is listed here for reference. }
\label{tab:tm}
\begin{tabular}{llllllll}
\\
\hline
\hline
Source name  & Distance (kpc) &  $P$ (s) & $\dot{P}$ (s/s) &  $\dot{E}$ (erg/s)&  $\tau$ (yr) 
&  $B_{\rm s}$ (G) & $B_{\rm lc}$ (G)    \\
\hline
J0205+6449  &\ \ 3.20   & 0.0657 & 1.94$\times 10^{-13}$ &2.7$\times 10^{37}$&5.37$\times 10^{3}$&3.61$\times 10^{12}$&1.19$\times 10^{5}$ \\
J0358+5413  &\ \ 1.00   & 0.1564 & 4.40$\times 10^{-15}$&4.5$\times 10^{34}$&5.64$\times 10^{5}$&8.39$\times 10^{11}$&2.06$\times 10^{3}$ \\
J0534+2200  &\ \ 2.00   & 0.0330 & 4.21$\times 10^{-13}$&4.5$\times 10^{38}$&1.26$\times 10^{3}$  &3.79$\times 10^{12}$&9.55$\times 10^{5}$\\
J0537-6910  &\ \ 49.70  & 0.0161 & 5.18$\times 10^{-14}$&4.9$\times 10^{38}$&4.93$\times 10^{3}$ &9.25$\times 10^{11}$&2.07$\times 10^{6}$ \\
J0540-6919 &\ \ 49.70   & 0.0505 & 4.79$\times 10^{-13}$&1.5$\times 10^{38}$&1.67$\times 10^{3}$&4.98$\times 10^{12}$&3.61$\times 10^{5}$ \\
J0633+1746  &\ \ 0.19   & 0.2371 & 1.10$\times 10^{-14}$&3.2$\times 10^{34}$&3.42$\times 10^{5}$&1.63$\times 10^{12}$&1.15$\times 10^{3}$ \\
J0659+1414  &\ \ 0.29    & 0.3849 & 5.49$\times 10^{-14}$&3.8$\times 10^{34}$&1.11$\times 10^{5}$&4.65$\times 10^{12}$&7.65$\times 10^{2}$  \\
J0835-4510  &\ \ 0.28     & 0.0893 & 1.25$\times 10^{-13}$&6.9$\times 10^{36}$&1.13$\times 10^{4}$&3.38$\times 10^{12}$&4.45$\times 10^{4}$\\
J1016-5857  &\ \ 3.16    & 0.1074 & 8.08$\times 10^{-14}$&2.6$\times 10^{36}$&2.10$\times 10^{4}$&2.98$\times 10^{12}$&2.26$\times 10^{4}$  \\
J1048-5832  &\ \ 2.90    & 0.1237 & 9.61$\times 10^{-14}$&2.0$\times 10^{36}$&2.04$\times 10^{4}$&3.49$\times 10^{12}$&1.73$\times 10^{4}$  \\
J1101-6101  &\ \ 7.00    & 0.0630 & 8.60$\times 10^{-15}$&1.4$\times 10^{36}$&1.16$\times 10^{5}$&7.42$\times 10^{11}$&2.81$\times 10^{4}$  \\
J1124-5916  &\ \ 5.00    & 0.1350 & 7.53$\times 10^{-13}$&1.2$\times 10^{37}$&2.85$\times 10^{3}$&1.02$\times 10^{13}$&3.85$\times 10^{4}$ \\
J1357-6429  &\ \ 3.10     & 0.1661 & 3.60$\times 10^{-13}$&3.1$\times 10^{36}$&7.31$\times 10^{3}$&7.83$\times 10^{12}$&1.60$\times 10^{4}$ \\
J1420-6048  &\ \ 5.63    & 0.0682 & 8.32$\times 10^{-14}$&1.0$\times 10^{37}$&1.30$\times 10^{4}$&2.41$\times 10^{12}$&7.13$\times 10^{4}$ \\
J1509-5850  &\ \ 3.37    & 0.0889 & 9.17$\times 10^{-15}$&5.1$\times 10^{35}$&1.54$\times 10^{5}$&9.14$\times 10^{11}$&1.22$\times 10^{4}$ \\
J1513-5908  &\ \ 4.40   & 0.1513 & 1.53$\times 10^{-12}$&1.7$\times 10^{37}$&1.57$\times 10^{3}$  &1.54$\times 10^{13}$&4.15$\times 10^{4}$ \\
J1617-5055  &\ \ 4.74    & 0.0694 & 1.35$\times 10^{-13}$&1.6$\times 10^{37}$&8.13$\times 10^{3}$&3.10$\times 10^{12}$&8.70$\times 10^{4}$ \\
J1709-4429  &\ \ 2.60    & 0.1020 & 9.30$\times 10^{-14}$&3.4$\times 10^{36}$&1.75$\times 10^{4}$&3.12$\times 10^{12}$&2.72$\times 10^{4}$  \\
J1718-3825  &\ \ 3.49    & 0.0747 & 1.32$\times 10^{-14}$&1.3$\times 10^{36}$&8.95$\times 10^{4}$&1.01$\times 10^{12}$&2.26$\times 10^{4}$  \\
J1747-2958  &\ \ 2.52    & 0.0988 & 6.13$\times 10^{-14}$&2.5$\times 10^{36}$&2.55$\times 10^{4}$&2.49$\times 10^{12}$&2.42$\times 10^{4}$  \\
J1801-2451  &\ \ 3.80     & 0.1249 & 1.28$\times 10^{-13}$&2.6$\times 10^{36}$&1.55$\times 10^{4}$&4.04$\times 10^{12}$&1.95$\times 10^{4}$  \\
J1803-2137  &\ \ 4.40     & 0.1337 & 1.34$\times 10^{-13}$&2.2$\times 10^{36}$&1.58$\times 10^{4}$&4.29$\times 10^{12}$&1.68$\times 10^{4}$  \\
J1809-1917  &\ \ 3.27    & 0.0827 & 2.55$\times 10^{-14}$&1.8$\times 10^{36}$&5.14$\times 10^{4}$&1.47$\times 10^{12}$&2.43$\times 10^{4}$  \\
J1811-1925  &\ \ 5.00     & 0.0646 & 4.40$\times 10^{-14}$&6.4$\times 10^{36}$&2.33$\times 10^{4}$&1.71$\times 10^{12}$&5.92$\times 10^{4}$ \\
J1813-1749  &\ \ 4.70    & 0.0447 & 1.27$\times 10^{-13}$&5.6$\times 10^{37}$&5.60$\times 10^{3}$&2.41$\times 10^{12}$&2.53$\times 10^{5}$ \\
J1826-1334  &\ \ 3.61    & 0.1010 & 7.53$\times 10^{-14}$&2.8$\times 10^{36}$&2.14$\times 10^{4}$&2.80$\times 10^{12}$&2.51$\times 10^{4}$ \\
J1833-1034  &\ \ 4.10     & 0.0619 & 2.02$\times 10^{-13}$&3.4$\times 10^{37}$&4.85$\times 10^{3}$&3.58$\times 10^{12}$&1.42$\times 10^{5}$ \\
J1838-0655  &\ \ 6.60    & 0.0705 & 4.93$\times 10^{-14}$&5.5$\times 10^{36}$&2.27$\times 10^{4}$&1.89$\times 10^{12}$&5.05$\times 10^{4}$  \\
J1856+0113  &\ \ 3.30    & 0.2674 & 2.08$\times 10^{-13}$&4.3$\times 10^{35}$&2.03$\times 10^{4}$&7.55$\times 10^{12}$&3.70$\times 10^{3}$  \\
J1930+1852  &\ \ 7.00   & 0.1369 & 7.51$\times 10^{-13}$&1.2$\times 10^{37}$&2.89$\times 10^{3}$ &1.03$\times 10^{13}$&3.75$\times 10^{4}$ \\
J1932+1059  &\ \ 0.31    & 0.2265 & 1.16$\times 10^{-15}$&3.9$\times 10^{33}$&3.10$\times 10^{6}$&5.18$\times 10^{11}$&4.18$\times 10^{2}$  \\
J1952+3252  &\ \ 3.00    & 0.0395 & 5.84$\times 10^{-15}$&3.7$\times 10^{36}$&1.07$\times 10^{5}$&4.86$\times 10^{11}$&7.38$\times 10^{4}$ \\
J2021+3651  &\ \ 10.51   & 0.1037 & 9.57$\times 10^{-14}$&3.4$\times 10^{36}$&1.72$\times 10^{4}$&3.19$\times 10^{12}$&2.68$\times 10^{4}$ \\
J2124-3358$^\dagger$  &\ \ 0.41   & 0.0049 & 7.27$\times 10^{-21}$&2.36$\times 10^{33}$&1.07$\times 10^{10}$&1.92$\times 10^{8}$&1.50$\times 10^{4}$ \\
J2225+6535  &\ \ 0.83    & 0.6825 & 9.66$\times 10^{-15}$&1.2$\times 10^{33}$&1.12$\times 10^{6}$&2.60$\times 10^{12}$&7.66$\times 10^{1}$ \\
J2229+6114  &\ \ 3.00    & 0.0516 & 7.83$\times 10^{-14}$&2.2$\times 10^{37}$&1.05$\times 10^{4}$&2.03$\times 10^{12}$&1.39$\times 10^{5}$ \\
\hline
\\
\end{tabular}
\end{table*}

\section{Correlations between spectral and timing properties}

\begin{figure}
\includegraphics[width=8cm]{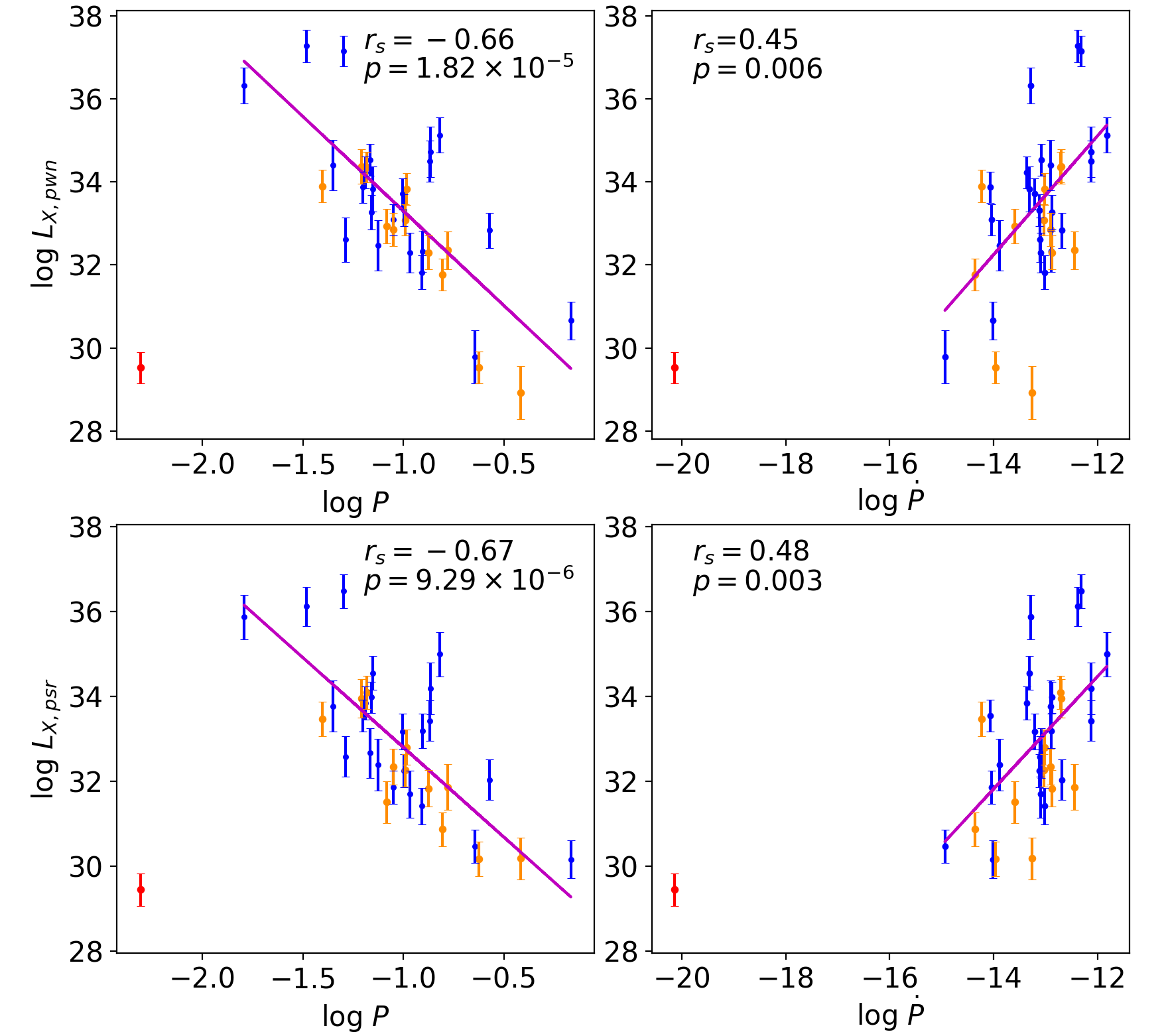}
\caption{Luminosity of PWNe ($\lxpwn$, upper panels) and of pulsars ($\lxpsr$, lower panels) versus  
pulsar period ($P$, left panels) and period time derivative ($\pdot$, right panels).
In the legend, $r_s$ is the Spearman rank-order correlation coefficient and $p$ is the corresponding random probability.
The 12 data points in orange are those with detected surface thermal emissions.
The purple lines are the best-fit linear function for the 23 data points in blue and 12 in orange. 
The red data point is PSR J2124-3358, which is a millisecond pulsar
and is not included
in determining $r_s$ and the linear fitting.
All the quantities are in gaussian units.
}
\label{fig:lpp}
\end{figure}

\begin{figure}
\includegraphics[width=8cm]{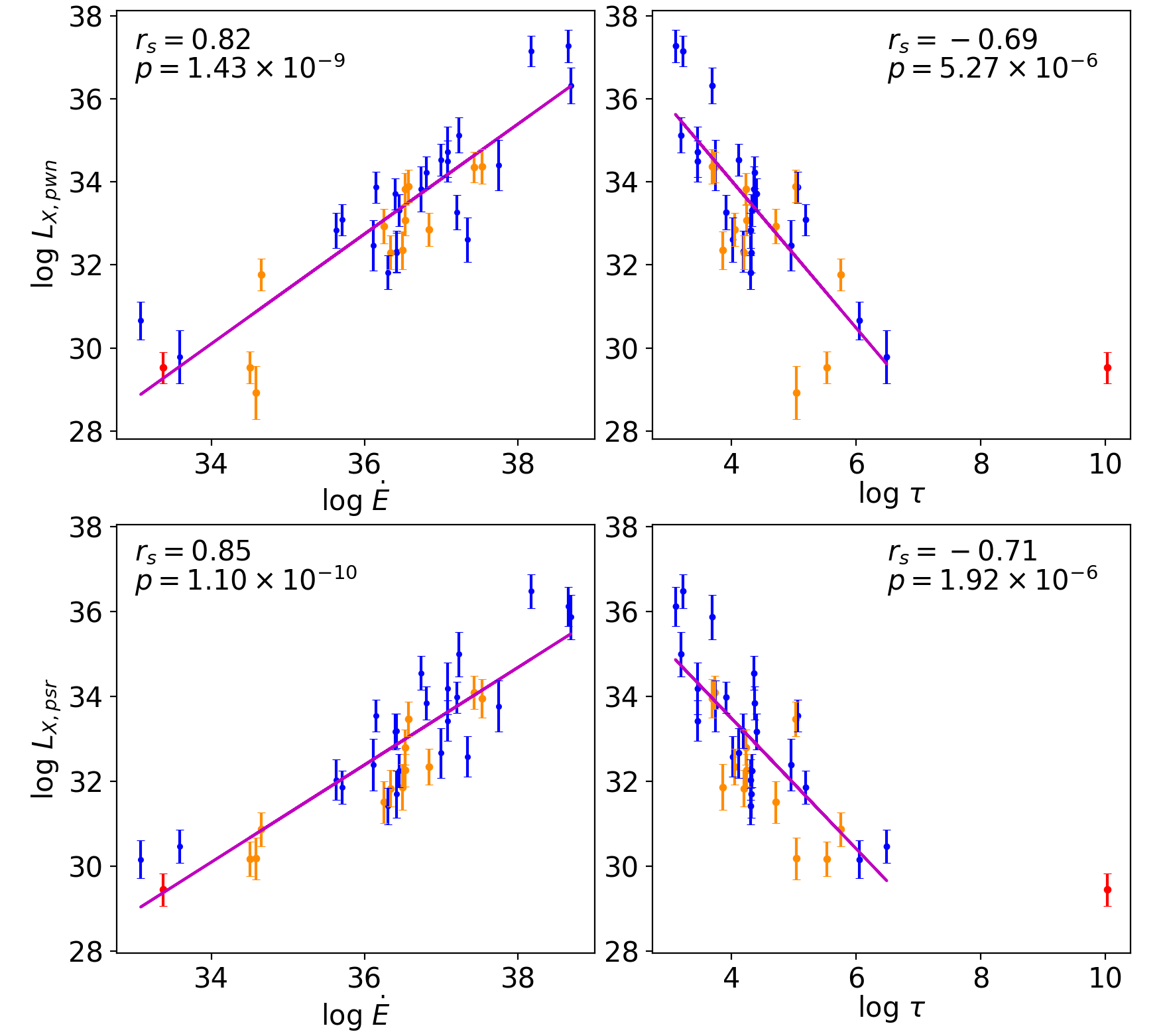}
\caption{The same as Fig. \ref{fig:lpp} but for $\lxpwn$ and $\lxpsr$ versus pulsar's spin-down power ($\edot$,  left panels)
and characteristic age ($\tau$, right panels).}
\label{fig:let}
\end{figure}

\begin{figure}
\includegraphics[width=8cm]{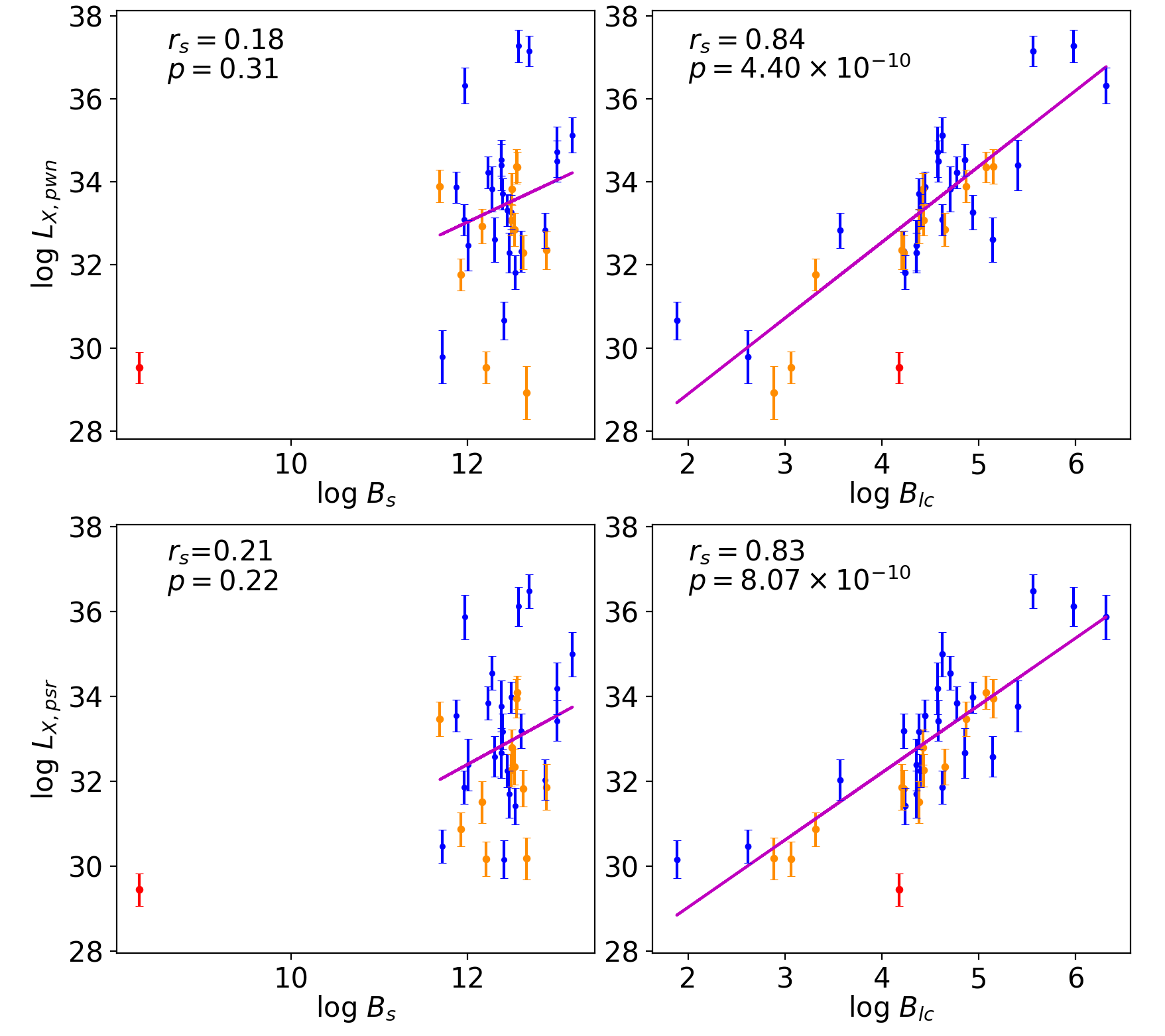}
\caption{The same as Fig. \ref{fig:lpp} but for $\lxpwn$ and $\lxpsr$ versus pulsar's dipole magnetic field strength at the surface 
 ($\bs$,  left panels)
and at the light cylinder ($\blc$, right panels).}
\label{fig:lbb}
\end{figure}

We first examine possible correlations between non-thermal X-ray luminosities ($\lxpwn, \lxpsr$) 
and pulsar timing properties.
The latter include period $P$, period time-derivative $\pdot$, spin-down power $\edot$, characteristic age $\tau$,
the dipole magnetic field strength at the stellar surface $\bs$ and at the light cylinder $\blc$.  
We adopt conventional values of the moment of inertia $I = 10^{45}$ g cm$^2$ and radius $R = 10$ km for
the pulsar, and have (all in gaussian units)
$\edot=I\Omega\dot{\Omega}=3.9\times 10^{46}P^{-3}\pdot$, 
$\tau=0.5P\pdot^{-1}$,
$\bs=3.2\times 10^{19}(P\pdot)^{0.5}$,
and
$\blc=\bs(R/R_{\rm lc})^3=2.9\times 10 ^8 P^{-2.5}\pdot^{0.5}$,
where $\Omega=2\pi/P$ and the light cylinder radius $R_{\rm lc}=c/\Omega$.
They are plotted in Fig.\ref{fig:lpp}, \ref{fig:let} and \ref{fig:lbb}.
We use the Spearman rank-order correlation coefficient $r_s$ to indicate the significance of possible correlations.
Each $r_s$ and its associated random probability $p$ are noted in the figures. 

As found in earlier literature, these luminosities are strongly correlated with $\edot$ (Fig. \ref{fig:let}).
We found, however, their correlations with $\blc$ are also equally strong (Fig. \ref{fig:lbb}).
The best fits to these correlations, using the least-square method to fit a linear function of the logarithms of these variables,  are
\beq
\lxpwn\propto\edot^{1.32\pm 0.13}\,\,\,\,\, (\chi^2_\nu=4.38)\,\,\, ,
\label{eq:lned}
\eeq
\beq
\lxpwn\propto\blc^{1.83\pm 0.19}\,\,\,\,\, (\chi^2_\nu=4.59)\,\,\, ,
\eeq
\beq
\lxpsr\propto\edot^{1.15\pm 0.11}\,\,\,\,\, (\chi^2_\nu=3.43)\,\,\, ,
\label{eq:lped}
\eeq
and
\beq
\lxpsr\propto\blc^{1.59\pm 0.17}\,\,\,\,\, (\chi^2_\nu=3.81)\,\,\, .
\label{eq:lpblc}
\eeq
The fitting uncertainties reported here and hereafter are all at a confidence level of 68\%.
The reduced chi-squares of these best fits are large and statistically not acceptable.
It reflects the fact that there is a huge scatter among data points in those corresponding figures,
although correlations seem strong.
We note that, in this work, best fits and chi-squares are only suggestive, because the quoted uncertainties
are dominated by distance uncertainties and are not derived rigorously with proper statistics. 
These unacceptable best fits are similar to what were found in earlier literature (e.g. \citet{possenti02,li08}).
The correlations of these luminosities with $P$ and with $\tau$ are weaker, and much weaker with $\pdot$. 
They are uncorrelated with $\bs$, as indicated by the Spearman's coefficient shown in Fig. \ref{fig:lbb}.

Although the magnitude of the fitting uncertainties to the best-fit power indices in Eq.(\ref{eq:lned}) and Eq.(\ref{eq:lped})
make the two power indices indistinguishable,
the trend that $\lxpwn$ depends on $\edot$ with a larger power index than $\lxpsr$ does,
as mentioned in Section 1, is still present. 
On the other hand,
$\edot$ and $\blc$ seem to play equally important roles in these correlations, when treated separately.
It is tempting that both factors should be taken into account together. 
We therefore checked whether a two-variable fitting gives a better description.
The best fits of $\log\lxpwn$ and $\log\lxpsr$ as a linear function of $\log\edot$ and $\log\blc$ are
\beq
\lxpwn\propto \edot^{0.83\pm 0.52}\blc^{0.71\pm 0.73}\,\,\,\,\, (\chi^2_\nu=4.38)
\label{eq:lnedblc}
\eeq
and
\beq
\lxpsr\propto \edot^{0.93\pm 0.48}\blc^{0.32\pm 0.68}\,\,\,\,\, (\chi^2_\nu=3.51)\,\,\,\, .
\label{eq:lpedblc}
\eeq
Based on the reported $\chi^2_\nu$ in these best fits and those in Eq.(\ref{eq:lned}) -- Eq.(\ref{eq:lpblc}),
it is clear that the fitting is not improved by invoking two variables.
Furthermore, the fitting uncertainties become too large to distinguish the dependence of $\lxpwn$ and $\lxpsr$ on
$\edot$ and $\blc$.

Since $\edot$ and $\blc$ are simply functions of $P$ and $\pdot$, we also fit
luminosity logarithms as a linear function of $\log P$ and $\log\pdot$ directly.
The best fits are 
\beq
\lxpwn\propto P^{-4.21\pm 0.54}\pdot^{1.10\pm 0.24}\,\,\,\,\, (\chi^2_\nu=4.46)
\eeq
and
\beq
\lxpsr\propto P^{-3.74\pm 0.49}\pdot^{1.00\pm 0.21}\,\,\,\,\, (\chi^2_\nu=3.47)\,\,\,\, .
\eeq
With $\edot\propto P^{-3}\pdot$ and $\blc\propto P^{-2.5}\pdot^{0.5}$, 
the above equations can be expressed in terms of $\edot$ and $\blc$ as the following:
\beq
\lxpwn\propto \edot^{0.65\pm 0.66}\blc^{0.91\pm 0.90}
\eeq
and
\beq
\lxpsr\propto \edot^{0.63\pm 0.58}\blc^{0.74 \pm 0.80}\,\,\,\, .
\eeq
In view of the large fitting uncertainties, the above two equations are not inconsistent with 
Eq.(\ref{eq:lnedblc}) and Eq.(\ref{eq:lpedblc}).

\begin{figure}
\includegraphics[width=8cm]{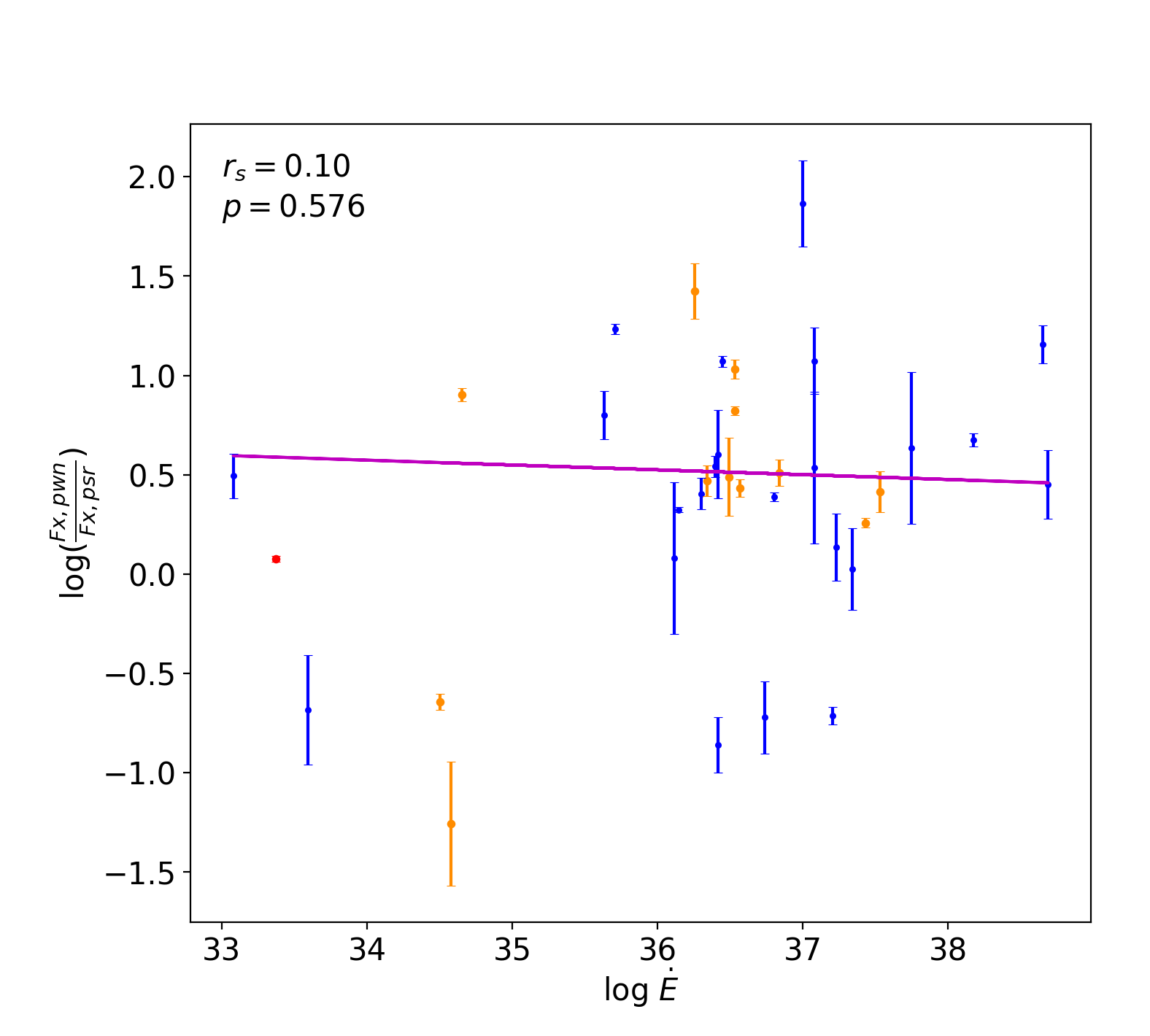}
\caption{X-ray flux ratio of nebulae to pulsars versus $\edot$. The purple line is the
linear best fit, which  has a slope of $-0.02\pm0.10$ but is unacceptable ($\chi^2_\nu=120$).
}
\label{fig:fred}
\end{figure}

The non-thermal luminosity of PWNe and of pulsars depends in a similar way 
on timing properties. 
Among the 35 neutron stars studied in this paper, 29 $\lxpwn$ are larger than the corresponding $\lxpsr$.
The ratio $\lxpwn/\lxpsr$, that is, the flux ratio $F_{\rm X,pwn}/F_{\rm X,psr}$, ranges from 0.06 to 73. 
This ratio, however, is not found to correlate with any timing properties.
One example, the relation between this ratio and $\edot$, is shown in Figure \ref{fig:fred}, 
in which one can see that their correlation is very weak. 
This ratio is not consistent with a constant either. 
The linear best fit shown in Figure \ref{fig:fred}, with a slope of $-0.02\pm0.10$ and a reduced chi-square equal to 120,
is statistically unacceptable.  

\begin{figure}
\includegraphics[width=8cm]{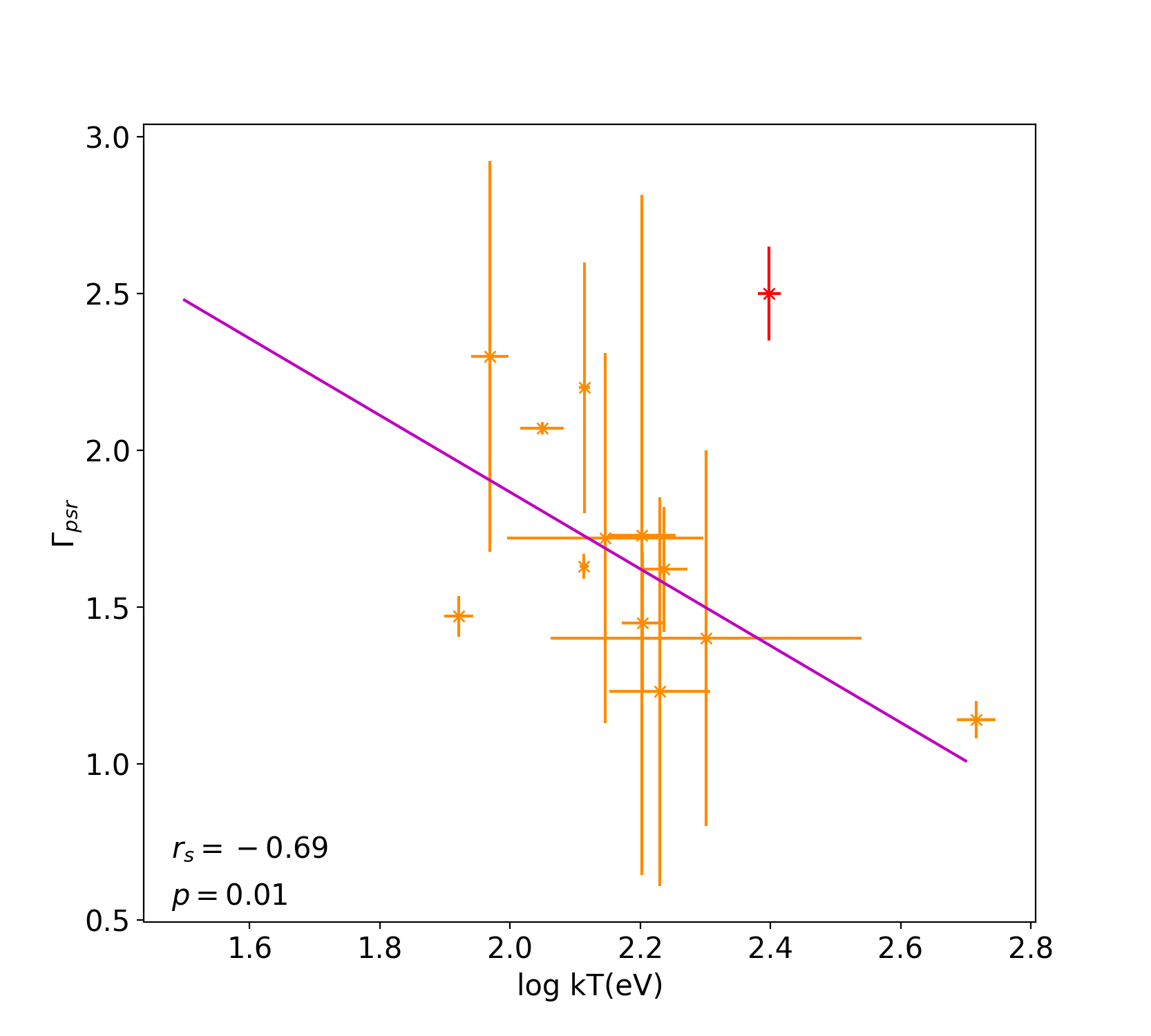}
      \caption{The same as Fig. \ref{fig:lpp} but with pulsar's photon indices ($\gmpsr$) versus surface temperatures ($T$). The orthogonal distance regression method is employed for the fitting.}
         \label{fig:gmt}
   \end{figure}

The photon index of the non-thermal X-ray power-law spectrum of the 35 pulsars we studied is not correlated
with any timing properties. The most significant Spearman's correlation coefficient is 0.29 (random probability 0.09) for
$\gmpsr$ versus $\log P$ and $-0.30$ versus $\log \blc$. 
The best fits to $\gmpsr$ as a linear function of logarithm of a single timing variable all yield 
a reduced chi-square of about 29, which is large, as expected from the very weak correlations. 
So is the fit to $\gmpsr$ as a linear function of two variables, $\log P$ and $\log\pdot$, together.
However, with the subset of 12 samples whose surface temperature estimate is observationally available (Table \ref{tab:sp}),
we found that the correlation between $\gmpsr$ and the surface temperature logarithm $\log T$ is modestly strong,
at a Spearman's correlation coefficient of $-0.69$ (random probability 0.01, Fig. \ref{fig:gmt}).
Its best-fit function is
\beq 
\gmpsr=4.32^{+0.66}_{-0.66}-1.23^{+0.31}_{-0.31}\log T \,\,\,\,\, (\chi^2_\nu=9.55) \,\,\,\, ,
\eeq
where, and hereafter, $T$ is in units of eV$/k$. The reduced chi-square $\chi^2_\nu$ is still statistically unacceptable.

The photon index of this 12-sample subset is not too different from the whole 35-sample set.
Its correlations with timing variables and best-fit reduced chi-squares are all similar to that of the whole sample, 
as described in the last paragraph.  
As revealed in Fig. \ref{fig:lpp} -- Fig. \ref{fig:lbb}, these 12 pulsars do not form a separate population.
When the surface temperature $T$ is taken into account, nonetheless, the fitting is significantly improved,
with reduced chi-squares $\chi^2_\nu$ decreasing from between 22 and 32 when fitting without $T$ to 
between 0.65 and 8.43 when $T$ is included. 
The most significant one is 
\beqary
\gmpsr&=&8.21^{+0.56}_{-0.56}-1.40^{+0.11}_{-0.11}\log T\nonumber \\
& & - 0.61^{+0.09}_{-0.09}\log P
+0.31^{+0.03}_{-0.03}\log \pdot  \nonumber \\
& & (\chi^2_\nu=0.65) \,\,\,\, .
\label{eq:gmtppraw}
\eeqary
Without $T$, the best fit of $\gmpsr$ as a linear function of $\log P$ and $\log\pdot$ is
$\gmpsr=5.07^{+1.89}_{-1.89} - 0.24^{+0.51}_{-0.51}\log P
+0.27^{+0.14}_{-0.14}\log \pdot$, with a reduced chi-square $\chi^2_\nu=25.12$ for 9 degrees of freedom.
The $F$ statistic for adding one more variable, i.e. $\log T$,
 is $F=\Delta\chi^2/\chi^2_\nu=(25.12\times 9 - 0.65\times 8)/0.65\approx 340$, corresponding to
a $p$ value way below $10^{-5}$, indicating
a very significant improvement of the fitting.
This equation suggests a possible dependence of $\gmpsr$ on $T$, $P$ and $\pdot$ as the following:
\beqary
\gmpsr &=& 8.21+\log(T^{-1.40}P^{-0.61}\pdot^{0.31})\nonumber \\
 &\approx & 8.21+\log(T^{-\frac{3}{2}}P^{-\frac{2}{3}}\pdot^{\frac{1}{3}}) \,\,\,\, .
 \label{eq:gmtpp}
 \eeqary
The combination of $P$ and $\pdot$ to certain powers can always be turned into a combination of 
any two timing variables. 
Unlike the case of $\lxpwn$ and $\lxpsr$, we do not see $\gmpsr$ correlate more with 
any particular timing variable. We therefore keep the factor
$P^{-\frac{2}{3}}\pdot^{\frac{1}{3}}$ here, awaiting more physical interpretations.

\section{Discussion}

The scatter of data point distribution in Fig. \ref{fig:lpp}--\ref{fig:gmt} is quite large, 
even for the cases with stronger correlations, such as $\lxpwn$ versus $\edot$ and $\blc$, 
and $\lxpsr$ versus $\edot$ and $\blc$.
This may indicate there are other dependences missing in these figures and may also
be due to huge uncertainties in the spectral properties of the samples that we collect.
These uncertainties include that from the original observations, 
from our choice of the nebula regions for PWNe with complex morphology, and, most importantly,
from the distance estimate. 
In addition, emissions from pulsars are probably viewing-geometry dependent. 
The luminosity $\lxpsr$ derived from the observed flux does not necessarily represent 
the true luminosity of the pulsar in question.
Emissions from PWNe are even more complicated.
They depend not only on properties of pulsar winds, but also on properties of the environments (e.g. \citet{kolb17}).
Nonetheless, some intriguing correlations between these luminosities and pulsars' timing variables do exist.

The correlation between X-ray luminosities and the spin-down power $\edot$ has been extensively studied
in the literature, as briefly reviewed in Section 1. 
With a larger sample than before, we obtain similar results.
Possible dependence on the magnetic field strength at the light cylinder, however, was somehow overlooked in earlier studies.
If pulsars' non-thermal X-ray emission comes from a region close to the light cylinder, 
the magnetic field strength there will very likely affect the X-ray luminosity.
The nebula emission, presumably synchrotron radiation in nature, depends on energetics of charged particles
in the pulsar winds and on the magnetic field strength around the termination shock.
If these charged particles are accelerated by magnetic reconnection, no matter along the way from the light cylinder to 
the termination shock or close to the termination shock, the field strength at the light cylinder may matter in some way.
It is therefore not surprising to see $\lxpwn$ and $\lxpsr$ have a strong correlation with $\blc$.
In fact we found this correlation is as strong as the one with $\edot$.
It suggests that these two timing variables together may describe luminosities better.
Unfortunately, large uncertainties in the fitting with currently available data hinder the progress towards this direction.
Nonetheless, because
synchrotron radiation luminosity depends on the magnetic field strength to the second power
and on the energy distribution of emitting particles,
the power indices of $\blc$ in Eq.(\ref{eq:lnedblc}) and Eq.(\ref{eq:lpedblc}), which are not close to 2,
indicate that $\blc$ may play some role in
determining energetics of those emitting particles, together with $\edot$.


The flux ratio, $\Fxpwn/\Fxpsr$, is not found to correlate with any timing properties.
It  indicates that, within the current uncertainty level, 
$\lxpwn$ and $\lxpsr$ have a very similar
dependence on pulsars' timing properties.
This can also be seen from all the fitting results shown in Eqs.(\ref{eq:lned}) -- (\ref{eq:lpedblc}).
It is, however, intriguing that $\lxpwn$ seems to depend on $\edot$ or $\blc$ with a larger power index than $\lxpsr$
when considering $\edot$ and $\blc$ dependence separately. 
We note that this trend is not fully supported by the current statistics because of the large uncertainty,
but it also appears  in all the earlier literature as described in Section 1.

Another major discovery of this work is the dependence of $\gmpsr$ on $\log T$, $\log P$ and $\log \pdot$.
While the best fit of $\gmpsr$ as a function of $\log P$ and $\log \pdot$ has  $\chi^2_\nu=25.12$,
by adding one more variable, $\log T$, the best-fit  $\chi^2_\nu$ becomes 0.65.
One can see that, from the $F$-test, it is a very significant improvement.
From the fitting result, we suggest  that $\gmpsr$ is equal to a constant plus 
$\log(T^{-\frac{3}{2}}P^{-\frac{2}{3}}\pdot^{\frac{1}{3}})$ (Eq.(\ref{eq:gmtpp})).
If we consider this is synchrotron radiation emitted by a population of  electron-positron pairs
with a power-law energy distribution $N_E\propto E^{-\alpha}$, we have 
$\alpha=2\gmpsr -1$. 
Therefore $\alpha$  is equal to a constant plus 
$\log(T^{-3}P^{-\frac{4}{3}}\pdot^{\frac{2}{3}})$.

Non-thermal X-ray emissions from pulsars are proposed to emit by secondary pairs in the pulsar magnetosphere
away from the stellar surface (e.g. \citet{takata08,harding15}).
The production of these pairs can be just above the polar cap via one-photon pair production 
\citep{daugherty82} in the strong magnetic field close to pulsar's magnetic pole, 
or in the outer gap via two-photon pair production \citep{cheng86} in a thermal X-ray photon bath originating
from the stellar surface.
In the former scenario, curvature gamma-ray photons emitted by accelerated primary particles make pairs through
interactions with the ambient strong magnetic field. 
The energy of the created pairs is related to the energy of primary particles, which are accelerated by 
electric fields parallel to magnetic fields in the polar gap and at the same time may lose energy to surrounding
thermal X-ray photons coming from the stellar surface through inverse Compton scattering
\citep{chang95,sturner95,harding98,timokhin19}. 
It is then possible that $T$ plays a certain role in determining the energy distribution of the secondary pairs.
In the latter scenario, since pairs are created in the outer gap through photon-photon collisions between
curvature photons from primary particles and thermal X-ray photons from the stellar surface, it is obvious that
surface temperature $T$ plays a role in determining the energy distribution of the created secondary pairs,
which, as well as for the case above the polar cap, of course also depends on energetics of primary particles.
It in turn brings in the dependence on $P$ and $\pdot$. 
Detailed modeling work is required to understand all these dependences. 

The millisecond pulsar PSR J2124-3358 is an outlier in the $P$ and $\pdot$ distributions as shown in
Fig. \ref{fig:lpp}, 
but it fits pretty well into the 35 samples in the $\lxpwn$-$\edot$, $\lxpsr$-$\edot$ and $\lxpwn$-$\lxpsr$
relations and to a lower degree in the $\lxpwn$-$\blc$ and $\lxpsr$-$\blc$ relations.
It seems to indicate that millisecond pulsars share the same energy conversion physics with other ordinary pulsars, 
despite of their much shorter rotation periods and much larger characteristic ages.
However, in the $\gmpsr$-$T$ distribution (Fig. \ref{fig:gmt}), 
one can see that its non-thermal power-law spectrum is softer than the other 12 ordinary pulsars, 
even though its surface temperature is relatively high.
Furthermore, if we include PSR J2124-3358 in the fitting of $\gmpsr$ as a linear function of $\log T$, $\log P$ and
$\log\pdot$, the best fit has $\chi^2_\nu=8.44$, 
much larger than 0.65 obtained when only the 12 ordinary pulsars are included (Eq.(\ref{eq:gmtppraw})).
This pulsar is not in the fundamental plane defined by Eq.(\ref{eq:gmtpp}).
  
We note that the spin-down power $\edot$ of PSR J2124-3358 is small.
The maximum available potential drop to accelerate charged particles in pulsar's magnetosphere is proportional to
$\edot^{0.5}$ (see, e.g., \citet{goldreich69,ruderman75}). 
If the electric field to accelerate charged particles is not strong enough, 
the electrical acceleration may be balanced by the energy loss due to inverse Compton scattering
against thermal photons \citep{chang95}.   
One possibility is that there are two fundamental planes in the \{$\gmpsr,T,P,\pdot$\} space.
One is for those pulsars whose $\edot$ is large enough so that primary particles can reach high energies,
and the other is for those with smaller $\edot$ whose primary particles can only reach a saturation energy, which is relatively lower.
The energy distribution of radiating pairs is of course dependent on the energetics of primary particles.
To verify this conjecture, more samples with smaller $\edot$ are desired.


\section*{Acknowledgements}
This work is supported by the Ministry of Science and Technology (MOST) of the Republic of China (Taiwan) under grants
MOST 108-2112-M-007-003 and MOST 109-2112-M-007-009.



\section*{Data availability}
The data underlying this article are available in the article and in the quoted references.


\label{lastpage}
\end{document}